\documentstyle[12pt]{article}

\small
\begin{document}

\title{
{\bf The breathing mode in extended Skyrme model. }
}
\author{ {\bf Abdellatif Abada}\thanks {email : "ABADA@ipncls.in2p3.fr"}
{\bf ~and Houari Merabet }\thanks {email : "MERABET@ipncls.in2p3.fr"}
\\
{\normalsize
Division de Physique Th\'eorique,
\thanks{Unit\'e de Recherche des Universit\'es Paris XI
et Paris VI associ\'ee au C.N.R.S } }\\
{\normalsize	Institut de Physique Nucl\'eaire,} \\
{\normalsize	F-91406 , Orsay Cedex, France }
}
\date{}
\maketitle
\begin{abstract}
We study an extended Skyrme model which includes fourth and sixth-order
terms. We explore some static properties like the
$\Delta$-nucleon mass splitting and investigate the Skyrmion
breathing mode in the framework of the linear response theory.
We find that the monopole response function has a pronounced peak located at
$\sim$ 400 MeV, which we identify to the Roper resonance $N(1440)$.
As compared to the standard one, the extended Skyrme model provides
a more accurate description of baryon properties.
\end{abstract}


\vskip 1cm
{\it Submitted to Phys. Rev. D }

\vskip .5cm
IPNO/TH 93-08 \hfill{March 1993}

\newpage
\begin{center}  {{\large {\bf INTRODUCTION }} } \end{center}

At low energy the QCD running coupling constant $\alpha_s$ becomes large
which renders the standard perturbation theory inapplicable.
In order to describe hadronic physics at low energy, effective theories
including the main features of QCD (e.g. chiral  symmetry)
have therefore been proposed. These theories carry the label "effective"
because their degrees of freedom are the hadronic observable instead
of the fundamental constituents, quarks and gluons, which are confined.
Ten years before the advent of QCD, Skyrme proposed a model \cite{Sk} for
hadronic physics which involves only meson fields (pions), and where baryons
emerge as topological solitons.
This model is now recognized as the simplest chiral
realization of QCD at low energy and large $N_c$ \cite {Wi}. In the following
we will refer to this model as the minimal Skyrme model.

The minimal Skyrme model has been studied extensively to describe static
properties of baryons [3-7]
as well as dynamical properties, in
particular the simplest vibrational excitation of the Skyrmion : the breathing
mode [8-15].
Concerning the static properties,
the predictions of the model are generally within 30$\%$ of experimental
values when the parameters are adjusted to fit the nucleon and the $\Delta$
masses. In this paper we investigate whether
this discrepancy is due to the use of the minimal Skyrme
model.  One motivation for this study is that it is easy to show
that the second and the fourth order terms in the pion field derivatives have
the same contribution to the soliton mass so that further terms should be
added. Moreover, in taking into account the Casimir energy
of the Skyrmion, the authors of Refs. \cite {MoDi,Mo} have found that
the nucleon mass is lowered to a value between 0 and 400 MeV which is much too
small ! They also proved that the $O(N_c^1)$ and $O(N_c^0)$ mass
contributions are of same order. Thus the large $N_c$ expansion is a poor
approximation within this minimal Skyrme model.

In this article we mainly focus on the breathing mode. One problem regarding
this mode in the minimal Skyrme model is that it does not show up in analysis
based on phase shifts \cite{Za,Br}, whereas when other calculations
found it \cite{Haj,Hay,Wa,Bie,Mat,Aba}, its excitation energy
stands between 200 and 300 MeV which is too small compared to its experimental
value of 500 MeV.

One possible way to
circumvent these difficulties is to improve this model by adding higher
order terms in derivatives in the corresponding chiral Lagrangian as the
sixth order term generated by $\omega$-meson exchange \cite{Mo}
(first proposed in Ref. \cite{Jac}).

This extended model has already been used to describe the
Roper resonance. Kaulfuss and Meissner \cite {Kau} used both scaling
approximation and semiclassical quantization.
They found a resonance at an excitation energy of $\sim$ 480 MeV.
However, these authors used values for the parameters of the model
which are in conflict with those determined by chiral perturbation theory
\cite{Gas,Ri} and, moreover, changed the sign of the so-called Skyrme term.
Schwesinger and Weigel \cite{Sch} have also used phase shifts analysis
within this extended Skyrme model. However, they did not find the Roper
resonance.

In this work, we describe this low-lying monopole resonance within the same
extended Skyrme model of Refs. \cite{Mo,Sch}
using the linear response theory. This method is more transparent and has
already been shown to be powerful in
different domains such as giant resonances in nuclear physics \cite{Rin} or
nucleon polarizabilities in hedgehog models in hadronic physic \cite{Bro}.
In a previous article \cite{Aba} we demonstrated the practicability of this
approach. However this calculation was limited to the minimal Skyrme model
only.

The present article is organized as follows. In Sec. I we introduce the
extended Skyrme model and define our notation. In
Sec. II we review the linear response approach of Ref. \cite{Aba} and
specialize to the extended model considered here.
We finally discuss our results concerning some static properties of the
soliton and the Roper resonance in Sec. III.

\renewcommand{\theequation}{\arabic{section}.\arabic{equation}}
\setcounter{equation}{0}
\section {The extended Skyrme model}

The simplest chiral effective Lagrangian proposed by Skyrme \cite{Sk} reads :
\begin{equation} \label{1}
{\cal L}_{SK} = {\displaystyle \left[ \frac {F_{\pi}^2 }{16} \right]
{}~\hbox {Tr} ~(\partial_{\mu} U \partial^{\mu} U^{+})
+ \left[ \frac {1}{32 e^2} \right]
{}~\hbox {Tr} ~\{~\left[ (\partial_{\mu} U) U^{+} , (\partial_{\nu} U) U^{+}
\right]^2  ~\} }
\end{equation}
where $U$ is an SU(2) matrix parametrized by the (Goldstone) pion fields
$\pi_a$, normalized to the pion decay constant $F_{\pi}$ :
\begin{equation} \label{1a}
U = {\displaystyle \exp[~2i \vec {\pi}. \vec {\tau} / F_{\pi} ~] }~,
\end{equation}
the $\tau_{a}$'s being the usual Pauli matrices.
The first term in the Lagrangian (\ref{1}) corresponds to the well known
nonlinear $\sigma$ model and the second one, which is of fourth-order in
powers of the derivatives of the pion field, was introduced by Skyrme to
stabilize the soliton. It is generally referred to as the Skyrme term.
It can be derived from a local approximation of an effective
model with $\rho$-mesons \cite{Ik}. Similarly, a term of order six can be
generated from $\omega$-meson exchange \cite{Jac,Ad2}. It reads :
\begin{equation} \label{2}
{\cal L}_6 = {\displaystyle -\frac {1}{2} \frac{\beta_{\omega}^2}{m_{\omega}^2}
B_{\mu} B^{\mu} } \end{equation}
where the anomalous baryon current $B_{\mu}$ is given by
\begin{equation} \label{3}
B_{\mu} = {\displaystyle \frac{1}{24 \pi^2}
\epsilon ^{\mu \nu \alpha \beta} \hbox {Tr} \left \{~
(\partial_{\nu} U) U^+ (\partial_{\alpha} U) U^+ (\partial_{\beta} U) U^+
{}~\right \} } .\end{equation}
The two new constants appearing in Eq. (\ref{2}) are the $\omega$-meson mass
$m_{\omega}$ and the parameter $\beta_{\omega}$ which can be related
to the $\omega \to \pi \gamma $ width \cite{La}.

In Ref. \cite {Mo} it was shown that the effect of the Casimir energy,
including the term ${\cal L}_6$ in the Lagrangian (\ref{1}), is
to lower the nucleon mass by about 500 MeV, so that one obtains a more
satisfactory value of $\sim $1 GeV.
Anticipating on the last section, let us observe
that the axial vector coupling constant $g_A$ is found to be 1.24 instead of
the Skyrme model prediction 0.34.
So there are good reasons to think that the extension
of the Skyrme model which consist in adding the sixth order term (\ref{2}) to
the Lagrangian (\ref{1}) is a good approximation to describe the baryonic
sector.

There are of course other sixth order terms in the
pion field derivatives such as the contributions of
the $\rho$ and scalar mesons. Nevertheless, a
Lagrangian containing all possible sixth order terms
leads in general to an Euler-Lagrange differential equation of
order higher than two. Consequently, one is not sure to find a soliton type
solution. This question is still open up to now since the parameters
which correspond to these terms have not yet been determined.
Fortunately, the term ${\cal L}_6$ in Eq. (\ref{2}) has the noteworthy
property (as the Skyrme term) of being quadratic in the time derivatives and
to lead to an equation of motion of second order as we will see below (see Eq.
(\ref{6})). For this reason we consider only this term in this work.

Our starting point is then the following Lagrangian :
\begin{equation} \label{4}
{\cal L} = {\cal L}_{SK} + {\cal L}_6 + {\displaystyle
\frac{1}{16} F_{\pi}^2 m_{\pi}^2 \hbox{Tr} \left(~U + U^+ -2 ~\right) ~~}.
\end{equation}
The first and second term in Eq. (\ref{4}) have been discussed above.
The last term which is proportional to the square of the pion mass $m_{\pi}$
implements a small explicit breaking of chiral symmetry.
Assuming the hedgehog ansatz for the pion field
$\vec{ \pi} ({\bf r},t) = F_{\pi} F(r,t) ~\hat{ {\bf r}}/2$ ,
(see Eq. (\ref{1a})) the Lagrangian density (\ref{4}) becomes
\begin{equation} \label{5} \begin{array}{ll}
{\cal L} = &{\displaystyle
\left[~ \frac {F_{\pi}^2 }{8} + \frac {\sin^2 F}{e^2 r^2} +
\frac{\beta_{\omega}^2}{8 m_{\omega}^2 \pi^4}
\left( \frac{ \sin F}{r} \right)^4 ~\right]
\partial_{\mu} F \partial^{\mu} F }\\
&{\displaystyle
- \left( \frac {\sin F}{2 r} \right)^2
\left[ F_{\pi}^2 + 2 \frac {\sin^2 F}{e^2 r^2} \right]
+ \frac{1}{4} F_{\pi}^2 m_{\pi}^2 (\cos F - 1) } ~.
\end{array} ~ \end{equation}
The corresponding classical Euler-Lagrange equation reads
\begin{equation} \label{6}
{\displaystyle
g^2 \ddot F + \alpha \dot F^2 = \left( g^2 F' \right)'
- \left( \theta + \alpha F'^2 \right) } ~~,\end{equation}
where the time-dependent functions $g, \alpha$ and $\theta$ are, respectively,
\begin{equation} \label{7} \begin{array} {ll}
g(r,t) = &{\displaystyle
\left[ \frac {(eF_{\pi})^2}{4} r^2 + 2 \sin^2 F +
\frac{\beta_{\omega}^2~e^2}{4 m_{\omega}^2 ~\pi^4}\frac{ \sin^4 F}{r^2}
\right]^{\frac {1}{2}} } ~~,\\
\alpha (r,t) = &{\displaystyle
\left[ 1 + \frac{\beta_{\omega}^2~e^2}{4 m_{\omega}^2 ~\pi^4}
\left( \frac{ \sin F}{r} \right)^2  \right] \sin (2F) } ~~,\\
\theta (r,t) = &{\displaystyle
\left[ \frac {(eF_{\pi})^2}{4} r^2 + \sin^2 F \right] \frac {\sin (2F)}
{r^2} + \frac{1}{4} (eF_{\pi})^2 m_{\pi}^2 ~r^2 \sin F  }
{}~~. \end{array} \end{equation}
Primes and dots in Eq. (\ref{6}) indicate radial coordinate differentiations
and time differentiations respectively.
 In order to ensure that the baryon number is equal to one, the equation
of motion (\ref{6}) has to be solved with the following conditions :
\begin{equation} \label{8}
F(0,t) ~=~ \pi ~~~,~~~  F(\infty,t) ~=~ 0 ~~
\end{equation}
which are sufficient for a differential equation of second order.

\setcounter{equation}{0}
\section { Linear response analysis}

In order to describe the low-lying monopole vibrations of the Skyrmion within
the extended Skyrme model (\ref{4}) we explore the
response of the Skyrmion to an external infinitesimal monopole field
with a frequency $\Omega$. The response function is determined from the
evolution of the isoscalar mean square radius of the Skyrmion with respect to
the frequency $\Omega$.

An external time-dependent monopole field corresponds to the addition of the
following term to the Lagrangian density (\ref{5})
\begin{equation} \label{9}
{\cal L}_{\hbox {int}} = eF_{\pi}^3 r^2 B^0(r,t)
{}~\epsilon \sin(\Omega t) \exp(\eta t) \end{equation}
where $B^0(r,t)$ is the time-component of the baryon current (\ref{3})
$$B^0(r,t)  = {\displaystyle - \frac {1}{2\pi^2}
\frac {\sin^2(F)}{r^2} \frac {\partial F}{\partial r} } $$
and $\eta$ a vanishingly small positive number.
Adding this interaction term (\ref{9}) to the Lagrangian density
(\ref{5}), the new corresponding Euler-Lagrange equation reads
\begin{equation} \label{10}
{\displaystyle
g^2 \ddot F + \alpha \dot F^2 = \left( g^2 F' \right)'
- \left( \theta + \alpha F'^2 \right)
+ \epsilon \frac {(eF_{\pi})^3 r}{\pi^2} \sin^2 F \sin(\Omega t)
\exp(\eta t) }  \end{equation}
where $g, \alpha$ and $\theta$ have been defined in the previous section.
This last equation is to be solved with the boundary conditions
$ F(t=-\infty,r) = F_s(r)$ and $ \dot F(t=-\infty,r) = 0 $ where $F_s(r)$ is
the static solution of Eq. (\ref{6}).

Because the term (\ref{9}) is weak (in the domain $t \in [-\infty, 0]$ ),
it introduces small changes of the classical solution. We thus can treat this
solution in a linear approximation.
To first order in $\epsilon$, $F(r,t)$ has the form
$$ F(r,t) = F_s(r) + \delta F(r,t)~ + \delta F^*(r,t)~, $$
where $\delta F(r,t) $ is linear in the field strength $\epsilon$. Moreover,
its time-dependence reads
$$ \delta F(r,t) = -i\frac {\epsilon}{2} R(r)
\exp(i(\Omega - i\eta)t )$$
Thus, equation (\ref{10}) becomes
\begin{equation} \label{11}
{\displaystyle
( (\Omega - i\eta)^2 - {\cal A}_s ) (g_s R) = - \frac
{(eF_{\pi})^3 }{\pi^2} \frac {r \sin^2 F_s}{g_s} }
\end{equation}
where the function $g_s$ and the operator ${\cal A}_s$ are, respectively,
\begin{equation} \label{12}
g_s(r) = {\displaystyle
\left[ \frac {(eF_{\pi})^2}{4} r^2 + 2 \sin^2 F_s +
\frac{\beta_{\omega}^2~e^2}{4 m_{\omega}^2 ~\pi^4}\frac{ \sin^4 F_s}{r^2}
\right]^{\frac {1}{2}} } \end{equation}
\begin{equation} \label{12a} \begin{array} {rr}
{\cal A}_s \equiv {\displaystyle -\frac {\hbox{d}^2}{\hbox{d} r^2} ~+~
\frac {g_s^{''}}{g_s} ~-~ \frac {2}{g_s^2}  \Bigg(~
\sin (2F_s) \left[ F^{''}_s +
\frac{\beta_{\omega}^2~e^2}{4 m_{\omega}^2 ~\pi^4}
\left( \frac{ \sin^2 F_s}{r^2} F_s' \right)' ~~\right] } \\
{\displaystyle + \cos (2F_s) \left[ F_s'^2 - \frac {(eF_{\pi})^2}{4}
- \frac{2}{r^2} \sin^2 F_s ~\right] } \\
{\displaystyle - \frac{ \sin^2 F_s}{r^2} \left[ 1 +
\frac{\beta_{\omega}^2~e^2}{4 m_{\omega}^2 ~\pi^4} ~F_s'^2 ~\right]
{}~- \frac {(eF_{\pi})^2}{8} m_{\pi}^2~r^2 \cos F_s  ~ \Bigg) }
\end{array} \end{equation}
The isoscalar mean square radius \cite{Ad1} is given by
$${\displaystyle
\langle r^2 \rangle = -\frac {2}{\pi }
\int_0^\infty  r^2 \sin^2(F) F' ~\hbox{d} r }.$$
Up to first order in $\epsilon$ it reads
$$
\langle r^2 \rangle = \langle r^2 \rangle_s -i\frac {\epsilon}{2}
(~f( \Omega) \exp(i(\Omega -i\eta) t) -
f^*( \Omega) \exp(-i(\Omega +i\eta) t) ~)
$$
where $\langle r^2 \rangle_s $ is the static mean
square radius \cite{Ad1} and $f$ the response function
\begin{equation} \label{555}
f( \Omega) = {\displaystyle
\frac{4}{\pi} \int_0^\infty r \frac {\sin^2 F_s}{g_s}
(g_s R) } ~\hbox{d} r .
\end{equation}
By using equation (\ref{11}), we can extract the following
spectral representation of the response function (\ref{555})
\begin{equation} \label{15}
{\displaystyle
f( \Omega) = -\frac {1}{\pi} \sum_n
\frac {| \langle \phi \vert \phi_n\rangle |^2 }
{ (\Omega - i\eta)^2 - \omega_n^2 } }\ ,\end{equation}
where the state $\phi$ is defined by
\begin{equation} \label{13}
{\displaystyle
\langle r\vert \phi \rangle = \frac {2 eF_{\pi} }{\pi}
\frac { r \sin^2 F_s}{g_s} } ,
\end{equation}
and the $\phi_n$ are the eigenstates of the operator ${\cal A}_s$ (see Eq.
(\ref{12a})), with the eigenvalues $\omega^2_n$, normalized according to
$${\displaystyle
\int_0^{\infty} \phi_n(r) \phi_m(r) eF_{\pi} \hbox {d}r = \delta_{nm} } ~~.
$$
In Eq. (\ref{15}) the limit $\eta \to 0^+ $ is, as usual \cite{Rin}, implicit
and corresponds to the boundary condition specified above.
The quantity of interest here is the imaginary part of the
response function, which is directly related to the distribution of
collective strength (see e.g. \cite{Rin} ). The energy at which the imaginary
part of the response function (\ref{15}) exhibits an unbound peak is
identified with the excitation energy of the Roper resonance.

\setcounter{equation}{0}
\section { Results and summary }

In this section we present our results for some static properties and for the
energy of the Roper resonance within the extended
Skyrme model (\ref{4}) and compare them to those of the minimal one.
First of all, because it is a meson theory, the parameters of the model have
to be fixed by fitting the low energy meson observables.
Concerning the pion decay constant, the pion and the $\omega$-mesons masses,
the experiment yields 186 MeV, 139.5 MeV and 782 MeV, respectively.
For the dimensionless parameter $e$, Riggenbach et al \cite{Ri},
in analysing the $K_{l4}$ decays, reduce by a factor
two the error bars on the chiral low-energy constants \cite{Gas}. As a
consequence, one finds $e = 7.1 \pm 1.2$ \cite{Mo} which is surprisingly
very close to the value $e=2\pi$ proposed by Skyrme a long time ago \cite{Sk}.
The last parameter $\beta_{\omega}$ is obtained by fitting the $\omega \to \pi
\gamma$ width, yielding $\beta_{\omega} = 9.3$ \cite{La}. In order to compare
the minimal and the extended Skyrme models we will consider two sets of
parameters :

[i]  $F_{\pi} = 186$ MeV , $e=7.1$ and $\beta_{\omega}= 0 $
(minimal Skyrme model ).

[ii]  $F_{\pi} = 186$ MeV , $e=7.1$ and $\beta_{\omega}= 9.3 $
(extended Skyrme model ).

In Tab. I we report the soliton mass, the isoscalar root mean square radius,
the axial vector coupling constant and the $\Delta$-nucleon mass splitting
for the different sets of parameters. In the third row we report the
experimental values \cite{data}.

Concerning the soliton mass we find 972 MeV and 1563 MeV in the case [i] and
[ii], respectively. We have to subtract $\sim 1$ GeV (Casimir energy) in the
case of the minimal model (case [i]) which leads to a value of $\sim$ 0 MeV
\cite {Mo} ! In the extended Skyrme model (case [ii]) we must subtract
$\sim 500$ MeV \cite{Mo} which leads to a reasonable value of $\sim 1$ GeV.
Regarding the $\Delta$-nucleon mass splitting, which is not affected by the
Casimir effect, we find 1192 MeV in the case [i] which is not a realistic
value compared to the experimental one 290 MeV. However, in the case [ii]
we obtain 227 MeV which is more acceptable.
Furthermore, the axial vector coupling constant $g_A$ is found to be 0.34, in
the case [i], instead of the experimental value 1.23 which is very close to the
value predicted by the improved Skyrme model (case [ii]). Of course,
one has to account for the Casimir effect in the calculation of $g_A$.
However, the corrections seem to be small within this extended Skyrme model
\cite{Mo}.
For reference we plot in Fig. 1 the chiral function $F_s$ solution of the
static Euler-Lagrange equation (\ref{6}) for the minimal and extended model.

For the breathing mode of the Skyrmion, we display in Fig. 2 the
imaginary part of the response function (\ref{15}) for the two sets of
parameters. In both cases we find an unbound peak. In case [i] we find this
peak to be broad and located at 500 MeV.
Nevertheless, this result cannot be reliable
since the other predicted properties are unrealistic (see Tab.1).
Contrary, in the case of the extended Skyrme model, the response function
exhibits a pronounced sharp peak located at
$\sim$ 400 MeV which we identify to the Roper resonance. This value
corresponds to the Roper-nucleon mass splitting and, consequently, is not
sensitive to the Casimir effect.

As mentioned in the introduction, the authors of Ref. \cite{Sch} found no
trace of the Roper resonance within the same extended
Skyrme model. So it is natural to check if this result
is due to the parameters they chose. When we take their parameters
( $F_{\pi}$ = 142.4~MeV, $e$ = 9.92 and  $\beta_{\omega}$ = 13.6 )
to investigate the Roper resonance with our approach,
we find a sharp peak located at 320 MeV. This leads us to
think that the difficulty in finding the breathing mode resonance in
Ref. \cite{Sch} is due to the implementation of the phase shifts method.

To summarize the main results of our calculations, we plot in Fig. 3 the
energy spectrum of the $\Delta$ and Roper ($N^*$) resonances
for the different values of the parameters and compare them to the
experimental one.
We see obviously that the case [ii], which corresponds to the extended model,
is the closest to the experimental situation.
In order to obtain a better agreement one has to take
into account higher order terms in addition to ${\cal L}_6$
(see Eq. (\ref{2})) in the chiral Lagrangian.

Finally, the message that we want to transmit through this work is that one
should not restrict oneself to the standard Skyrme model \cite {Sk} (see Eq.
(\ref{1}) ) for the description of low energy hadron physics, but consider
extensions of this model including higher order terms in powers of the
derivatives of the pion field.
In this sense, the model considered here can be considered as a minimal
extension of the Skyrme Lagrangian.
This claim confirms the conclusions of Ref. \cite{Mo,Kala}.
A more realistic improvement consists of considering
effective Lagrangians which incorporate low mass mesons with finite mass
\cite{La,KaVin}.

\vskip 1cm
{\large {\bf Acknowledgments} }
We are grateful to D. Kalafatis, M. Knecht, J. F. Mathiot, B. Moussallam and
D. Vautherin for stimulating discussions and a critical reading of the
manuscript.
Special thanks are due to B. Moussallam for discussions regarding the Casimir
effects and for his help in numerical calculations of some static properties.
The Division de Physique Th\'eorique is " Unit\'e de Recherche des
Universit\'es Paris XI et Paris VI associ\'ee au CNRS".

\newpage
{\Large {\bf Table captions } }
\vskip 1cm

 {{\bf TABLE I.}} Are reported, for the two different combinations of the
parameters, the soliton mass $M_S$, the isoscalar root-mean-square radius
$r_0$, the axial vector coupling constant $g_A$, the $\Delta$-nucleon mass
splitting and the Roper-nucleon mass splitting $\hbar \Omega_{Roper}$.
The last line is derived from the data \cite{data}. $F_{\pi}$,
$m_{\pi}$ and $m_{\omega}$ are taken to their experimental values (see Sec.
III).

\begin{tabular}{ c c c c c c c}
\\ \hline\hline
$e$ & $\beta_{\omega}$ & $M_S$ & $r_0$ & $g_A$ &
$M_{\Delta} - M_N$ & $ \hbar \Omega_{Roper} $ \\
& & MeV & fm & & MeV & MeV \\
\hline
$7.1^{(a)}$ & $0$ & $972$ & $0.30$ & $0.34$ & $1192$ & $500$ \\
$7.1^{(a)}$ & $9.3^{(b)}$ & $1563$ & $0.61$ & $1.24$ & $227$ & $395$
\\
& & $939$ & $0.72$ & $1.23$ & $290$ & $500$ \\
\hline\hline
\end{tabular}

(a) from Ref. \cite{Ri} and (b) from Ref. \cite{La}

\vskip 1cm
{\Large {\bf Figure captions } }
\vskip 1cm

{\bf FIG. 1.} Static hedgehog solution $F_s(r)$ with experimental values of
$F_{\pi}$, $m_{\pi}$ and $m_{\omega}$ (see Sec. III).

{\bf FIG. 2.}  Imaginary part of the response function $f $~(fm$^2)$
versus the energy $ \Omega $ in MeV with experimental values of
$F_{\pi}$, $m_{\pi}$ and $m_{\omega}$ (see Sec. III).

{\bf FIG. 3.}  Energies splitting of the $\Delta$ and Roper $N^*$ resonances
with respect to the nucleon $N$ according to the cases [i] and [ii] of the
parameters (see Sec. III) and to the experiment \cite{data}.
All the energies are in MeV.


\newpage
\begin{thebibliography}{99}
\bibitem{Sk} T. H. R. Skyrme, Proc. Roy. Soc. {\bf A260}, 127 (1961).
\bibitem{Wi} E. Witten, Nucl. Phys. {\bf B223}, 422 and 433 (1983).
\bibitem{Ad1} G. S. Adkins, C. R. Nappi, E. Witten Nucl. Phys.
{\bf B228}, 552 (1983).
\bibitem{Jacks} A. Jackson, A. D. Jackson and V. Pasquier,
Nucl. Phys. {\bf A432}, 567 (1985)
\bibitem{La1} M. Lacombe, B. Loiseau, R. Vinh Mau and W. N. Cottingham, Phys.
Lett. {\bf 161B}, 31 (1985).
\bibitem{Vin} R. Vinh Mau, Prog. Part. Nucl. Phys. {\bf 20}, 221 (1988).
\bibitem{Me} U. G. Meissner Phys. Rep. {\bf 161}, 213 (1988).
\bibitem{Haj} Ch. Hajduk and B. Schwesinger Phys. Lett. {\bf 140B}, 172 (1984).
\bibitem{Hay} A. Hayashi and G. Holzwarth, Phys. Lett. {\bf 140B}, 175 (1984).
\bibitem{Za} I. Zahed, U. G. Meissner and U. B. Kaulfuss  Nuc. Phys.
{\bf A426}, 525 (1984).
\bibitem{Wa} H. Walliser and G. Eckart, Nucl. Phys. {\bf A429}, 514 (1984).
\bibitem{Br} J. D. Breit and C. R. Nappi, Phys. Rev. Lett. {\bf 53}, 889
(1984).
\bibitem{Bie} L. C. Biedenharn, Y. Dothan and M. Tarlini,
Phys. Rev. D{\bf 31}, 649 (1985).
\bibitem{Mat} M. P. Mattis and M. Karliner, Phys. Rev. D{\bf 31}, 2833 (1985).
\bibitem{Aba} A. Abada and D. Vautherin, Phys. Rev. D{\bf 46}, 3180 (1992).
\bibitem{MoDi} B. Moussallam and D. Kalafatis, Phys. Lett. B {\bf 272}, 196
(1991)
\bibitem{Mo} B. Moussallam preprint IPNO/TH 92-75 to appear in Ann. Phys. ;
private communication.
\bibitem{Jac} A. Jackson, A. D. Jackson, A. S. Goldhaber, G. E. Brown and L. C.
Castillejo, Phys. Lett. {\bf 154B}, 101 (1985).
\bibitem{Kau} U. B. Kaulfuss and U. G. Meissner, Phys. Lett. {\bf 154B}, 193
(1985).
\bibitem{Gas} J. Gasser and H. Leutwyler, Ann. Phys. {\bf 158}, 142 (1984).
\bibitem{Ri} C. Riggenbach, J. Gasser, J. F. Donoghue and B. R. Holstein, Phys.
Rev. D{\bf 43}, 127 (1991).
\bibitem{Sch} B. Schwesinger and H. Weigel, Nucl. Phys. {\bf A465}, 733 (1987).
\bibitem{Rin} P. Ring and P. Schuck, {\it The nuclear many body problem},
Springer (1980).
\bibitem{Bro} W. Broniowski and T. D. Cohen, Phys. Rev. D{\bf 47}, 299 and 313
(1993).
\bibitem{Ik} K. Iketani, Kyushu Univ. preprint, 84-HE-2 (1984).
\bibitem{Ad2} G. S. Adkins and C. R. Nappi, Phys. Lett. {\bf 137B}, 251 (1984).
\bibitem{La} M. Lacombe, B. Loiseau, R. Vinh Mau and W. N. Cottingham, Phys
Rev. D{\bf 38}, 1491 (1988).
\bibitem{data} Particle data booklet, {\it review of particle properties} Phys.
Rev. D {\bf 45}, Part 2 (june 1992).
\bibitem{Kala} D. Kalafatis, preprint IPNO/TH 92-76, to appear in the
proceedings of the workshop on QCD vacuum (World Scientific) Paris (1992).
\bibitem{KaVin} D. Kalafatis and R. Vinh Mau, Phys. Rev. D {\bf 46}, 3906
(1992).
\end{thebibliography}
\end{document}